\documentclass[final,twocolumn]{revtex4}
\usepackage{amssymb}
\usepackage{graphicx}
\usepackage{dcolumn}
\usepackage{amsmath}

\input{tcilatex}

\begin{document}

\title{Theoretical and Experimental Study of Stimulated and Cascaded Raman
Scattering in Ultra-high-Q Optical Microcavities }
\author{Tobias J. Kippenberg, Sean M. Spillane, Bumki Min, Kerry J. Vahala* 
\thanks{%
email: vahala@its.caltech.edu}}

\begin{abstract}
Stimulated Raman scattering (SRS) in ultra-high-Q surface-tension-induced
spherical and chip-based toroid microcavities is considered both
theoretically and experimentally. These microcavities are fabricated from
silica, exhibit small mode volume (typically 1000 $\mu m^{3}$) and possess
whispering-gallery type modes with long photon storage times (in the range
of 100 ns), significantly reducing the threshold for stimulated nonlinear
optical phenomena. Oscillation threshold levels of less than 100 $\mu $%
-Watts of launched fiber pump power, in microcavities with quality factors
of 100 million are observed. Using a steady state analysis of the
coupled-mode equations for the pump and Raman whispering-gallery modes, the
threshold, efficiencies and cascading properties of SRS in UHQ devices are
derived. The results are experimentally confirmed in the telecommunication
band (1550nm) using tapered optical fibers as highly efficient waveguide
coupling elements for both pumping and signal extraction. The device
performance dependence on coupling, quality factor and modal volume are
measured and found to be in good agreement with theory. This includes
analysis of the threshold and efficiency for cascaded Raman scattering. The
side-by-side study of nonlinear oscillation in both spherical microcavities
and toroid microcavities on-a-chip also allows for comparison of their
properties. In addition to the benefits of a wafer-scale geometry, including
integration with optical, electrical or mechanical functionality,
microtoroids on-a-chip exhibit single mode Raman oscillation over a wide
range of pump powers.
\end{abstract}

\affiliation{Department of Applied Physics, California Institute of Technology\\
Pasadena, CA 91125 }
\maketitle


\section{\protect\bigskip introduction}

Ultra-high-Q ( UHQ) surface-tension-induced microcavities (which we
subsequently refer to as STIMs) combine small modal volume with some of the
highest optical quality-factors (Q) to date of nearly 10 billion \cite%
{VernooyQ}, and are of interest for a variety of studies ranging from
fundamental physics such as cavity quantum electrodynamics \cite{VernooyQED}%
\cite{KimbleScripta}\cite{HarocheQED} to applied areas such as low threshold
and narrow linewidth lasers \cite{Sandoghdar}\cite{CaiOL}\cite{YangAPL},
nonlinear optical oscillators\cite{ChangBook}\cite{HydrogenSRSreference}, as
well as high-sensitivity transducers for biochemical sensing\cite{Vollmer}.
\ These silica microcavities feature whispering gallery type modes (WGMs)
and rely upon exquisite smoothness at the cavity dielectric boundary to
attain ultra-high-Q performance that is typically in excess of 100 million.
For nonlinear optical studies, strong resonant buildup of energy in
microscale volumes significantly reduces the threshold for nonlinear optical
effects to occur. This was recognized in the pioneering work of Chang \cite%
{ChangScience}\cite{ChangJOSA} and Campillo \cite{CampilloNLO}\cite%
{CampilloQED}\cite{CampilloOptComm}who observed and studied a variety of
nonlinear optical effects in ultra-high-Q liquid microdoplets. Their work
used free-space illumination to optically pump the microdroplets and thereby
induce Raman oscillation \cite{ChangJOSA}\cite{CampilloNLO}\cite%
{CampilloOptComm}, cascaded Raman scattering \cite{ChangJOSA} and Brillouin
scattering\cite{ChangBrillouin}. Silica UHQ STIMs provide a far more stable
and robust microcavity in comparison with liquid microdroplets. However,
despite numerous studies on these devices over the past decade \cite%
{Braginsky}\cite{Braunstein}\cite{Collot}\cite{Treussart}\cite%
{GorodetskyOptComm}\cite{GorodetskyOL}\cite{GorodetskyJOSA1} \cite%
{GorodetskyJOSA2}\cite{IlchenkoCoupling}\cite{IlchenkoMicrotorus}the
observation of nonlinear phenomena (beyond thermal effects) in these
devices, had been limited to one report on Kerr-induced wavelength shifts at
low temperatures \cite{Treussart}. The advent of pumping and signal
collection using fiber taper coupling methods \cite{Knight}\cite{SpillanePRL}%
\cite{CaiPTL}\cite{CaiPRL} proved an important turning point in access to
nonlinear phenomena in this important micro-cavity system. Fiber tapers
provide remarkably efficient coupling to and from the UHQ silica sphere
system \cite{SpillanePRL}\cite{CaiPRL}and enable direct access to the
technologically important optical fiber transport medium. The measured
fiber-coupled threshold for a variety of nonlinear phenomena in
taper-coupled, silica microspheres are lower than for any other nonlinear
oscillator reported to date. Silica microsphere Raman lasers with ultra-low
threshold levels of only 62 $\mu $-Watts \cite{SpillaneNature} have been
demonstrated. Compared to microdroplets these devices allow stable and long
term observation of nonlinear optical effects in microcavities. Cascaded
Raman lasing in these devices of up to 5 orders has also been observed \cite%
{Min} . \ The tapered optical fiber in these experiments functions to both
pump WGMs as well as to extract the nonlinear Raman fields. In addition, the
tapered-fiber coupling junction is highly ideal\cite{SpillanePRL}, making it
possible to strongly overcouple ultra-high-Q cavities with negligible
junction loss. This feature allows for the observation of very high \textit{%
internal} differential photon conversion efficiencies approaching unity.
Whereas microspheres are both compact and efficient nonlinear oscillators,
their fabrication properties lack the control and parallelism typical of
microfabrication techniques. Recently-developed ultra-high-Q toroid
microcavities on-a-chip \cite{Armani} provide a UHQ silica device with
performance equivalent to a microsphere. UHQ toroids have several advantages
over spheres including being wafer-scale devices that can be fabricated in
parallel as dense arrays or integrated with electronics or other optical
functionality. In this paper, we demonstrate and analyze nonlinear Raman
oscillation in both microsphere and microtoroid on-a-chip structures. In
addition to studying Raman oscillation in microcavities in detail, this work
allows us to compare the performance and properties of toroidal and
spherical microcavities. In addition to their fabrication and integration
advantages, it will be seen that microtoroids also have performance
advantages in comparison to microspheres. This includes a reduced number of
supported azimuthal modes, which allows observation of single-mode Raman
oscillation over a large range of pump powers\cite{KippenbergOLRaman} (of
critical importance in practical applications), as well lower Raman
threshold due to a reduced mode volume compared to a spherical cavity.

The paper is organized as follows. The first section will give a brief
introduction into the fabrication, coupling and optical properties of
ultra-high-Q toroid on-a-chip and spherical microcavities. The second
section will present a model for Raman lasing in a waveguide-coupled
whispering-gallery-mode resonator. There, we derive the expression for the
threshold and the efficiency of the conversion process. In the ultra-high-Q
regime resonances are easily split into doublets due to intermode coupling
of the degenerate clockwise and counterclockwise propagating whispering
gallery modes\cite{Weiss}. The effect of this intermode coupling on
stimulated Raman scattering is also considered in this section. The case of
cascaded Raman oscillation in which Raman signals serve to pump and generate
higher-order Raman waves is also treated in the section. In the fourth
section, the theoretical results are compared with experimental studies of
the dependence of Raman threshold on quality-factor, mode volume and
waveguide loading. Experimental results concerning power and efficiency
during cascaded operation are presented and compared to theory and found to
be in good agreement with theoretical predictions. Finally, a comparison of
microtoroid oscillation properties with those of microspheres is presented
in the fifth section.

\bigskip

\section{Ultra-high-Q surface-tension-induced microcavities}

\bigskip

Surface-tension-induced microcavities such as micro-droplets, exhibit a
superb, cavity surface finish (typically nanometer surface roughness\cite%
{VernooyQ}) leading to whispering gallery type modes with some of the
highest optical quality factors recorded to date. In the work presented
here, both spherical and toroid microcavities on-a-chip made from silica are
investigated. Both types of structures exhibit Q-factors in excess of 100
million. Briefly, the fabrication of both micro-sphere and micro-toroids
relies upon surface tension to induce collapse of a given silica preform
into the final cavity shape. Because the cavity fabrication involves a
temporary liquid state the surface finish of the final "solid" microcavity
is excellent. In the case of a microsphere, the preform used here is an
optical fiber tip which is heated and melted with a carbon-dioxide laser
(10.6 $\mu m$ wavelength). Surface tension causes the silica fiber tip to
contract and form a spherical microcavity, while the remainder of the fiber
stem serves as a holder for sphere-positioning. Light within the sphere is
confined near an equatorial plane by continuous total internal reflection at
the cavity interface. Microtoroids, on the other hand, use wafer
microfabrication techniques involving a combination of lithography and
etching combined with a final selective reflow process using a CO$_{2}$
laser. The details of this fabrication process are reported in reference 
\cite{Armani}. In the case of toroid microcavities the preform consists of a
microfabricated silica disk supported by a silicon pillar. Illumination of
the disk using a CO$_{2}$ laser induces selective reflow of the silica.
Surface tension causes the disk preform to collapse into a toroidal
periphery, thereby creating the resonant cavity. Figure 1 contains an
optical micrograph of a microsphere and a microtoroid on-a-chip.

\begin{figure}[tbp]
{\centerline{\includegraphics[width=7.6cm]{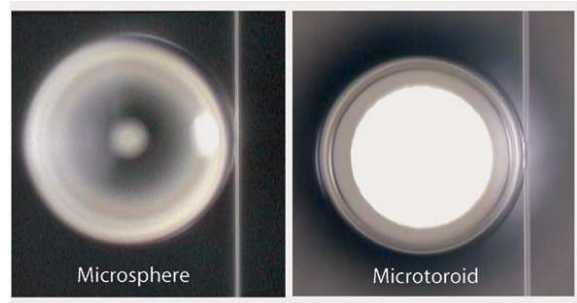}}}
\caption{Optical micrograph of a spherical (left) and toroid microcavity
(right) coupled to a tapered optical fiber.}
\end{figure}

Highly efficient evanescent coupling to these micro-cavities can be achieved
by use of tapered optical fibers \cite{CaiPRL}\cite{SpillanePRL}, which are
fabricated by melting and adiabatically tapering a standard,
telecommunication fiber until a waist diameter of approximately 1 to 2
microns is reached. When brought into proximity of the\ WGM resonator and
when the taper waist diameter is chosen to phase match to the WGMs,
efficient coupling both to and from the microcavity can be achieved.
Coupling can be described by the normalized coupling parameter ${%
K_{p}=\left( \frac{\tau _{0}}{\tau _{ex}}\right) }$ which describes the
ratio of intrinsic resonator lifetime\ $\tau _{0}$ to the external (coupling
related) lifetime $\tau _{ex}$. Following the standard conventions\cite{Haus}%
, undercoupling is denoted by $K<1$, overcoupling by $K>1$, and critical
coupling (the point of vanishing waveguide transmission) is denoted by $K=1.$
The ability to achieve strong overcoupling is important as it exemplifies
the very ideal nature\cite{SpillanePRL} of the taper-microcavity, coupling
junction. In addition to efficient excitation and extraction of optical
power, the fiber taper provides direct coupling to optical fiber, thereby
further facilitating laboratory measurements. In the case of microtoroids
on-a-chip the taper is also crucial as a means of "probing" the whispering
gallery devices which are within a few microns of the silicon wafer surface.

Using tapered optical fibers, resonator quality factor can be inferred from
either linewidth measurements or cavity ringdown experiments\cite{Armani}.
Figure 2 shows a cavity ringdown measurement on a 45-$\mu m$-diameter
microtoroid at 1550 nm, exhibiting a critically coupled Q of ${10^{8}}$. In
this measurement the WGM was excited on resonance, and the fiber-taper
adjusted to the critical point. The cavity lifetime can then be inferred by
gating-off the excitation laser, and recording the cavity decay signal. The
critically coupled Q (including the waveguide coupling contributions) was
100 million, and when correcting the Q for the waveguide loading, as
described in section III.B, an intrinsic quality factor of ${3.7\times 10^{8}%
}$ is inferred.

\begin{figure}[tbp]
{\centerline{\includegraphics[width=7.6cm]{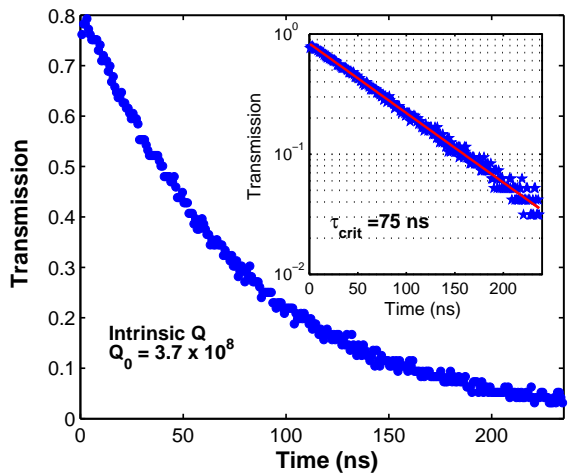}} }
\caption{Cavity rindown measurement of a 50-micron-diameter toroid
microcavity coupled to a tapered optical fiber. At t=0 the laser is gated
off, and (after an 8 ns fall-time) the transmission is entirely due to the
cavity decay field. Inset: A logarithmic plot to infer the decay time yields
75 ns at the critical point.\ }
\end{figure}

\section{Theoretical analysis of Raman Scattering in high-Q microcavites}

\subsection{1st Order Raman Scattering in microcavities}

Raman scattering in a waveguide coupled microcavity can be described
classically by using coupled mode equations for the pump and Raman fields
with nonlinear Raman coupling terms. Other nonlinear effects, which can
compete with Raman scattering, such as Four-wave-mixing or Brillouin
scattering are not considered in this analysis as the microcavity poses
stringent frequency matching constraints on these processes making their
observation difficult. In the case of Stimulated Brillouin Scattering, the
narrow gain bandwidth in the range of 100 MHZ, makes overlap of cavity modes
with the Brillouin gain spectrum unlikely and was not observed in this
work.\ In the case of parametric oscillation mediated by the
Kerr-nonlinearity (i.e. four-wave mixing), energy conservation requires a
triple resonance condition for signal, idler and pump mode, which can be
satisfied only under certain conditions\cite{KippenbergPRL}. These
additional constraints for oscillation based on the Kerr nonlinearity or
Brillouin gain, cause stimulated Raman scattering - which is intrinsically
phase-matched- to be the dominant microcavity nonlinear optical effect. For
simplification, we assume that the pump wavelength and the Raman wave are on
resonance and use the slowly varying envelope approximation. 
\begin{equation}
\begin{split}
\frac{dE_{p}}{dt}& =-\left( \frac{1}{2\tau _{ex}}+\frac{1}{2\tau _{0}}%
\right) _{p}E_{p}-\frac{\omega _{p}}{\omega _{R}}g_{R}^{c}(\omega
_{p,}\omega _{R})\cdot \left\vert E_{R}\right\vert ^{2}E_{p}+\kappa s \\
\frac{dE_{R}}{dt}& =-\left( \frac{1}{2\tau _{ex}}+\frac{1}{2\tau _{0}}%
\right) _{R}E_{R}+g_{R}^{c}(\omega _{p},\omega _{R})\cdot \left\vert
E_{P}\right\vert ^{2}E_{R}
\end{split}%
\end{equation}%
Here $E$ signifies the slowly-varying amplitude of the pump and Raman WGM
modes of the cavity and $s$ denotes the input wave. The excitation frequency
of the pump mode and resonant Raman mode is given by ${\omega _{R}}$ and ${%
\omega _{p}}$and ${\tau }$ is the total lifetime of photons in the
resonator, which is related to the quality factor by ${Q=\omega \cdot \tau }$%
. The coupling coefficient ${\kappa }$ denotes the coupling of the input
pump wave ${s}$ to the cavity whispering-gallery-mode ${E_{p}}$. The
relation ${\kappa =\sqrt{\frac{1}{\tau _{ex}}}}$ associates the coupling
coefficient with a corresponding lifetime, such that ${\frac{1}{\tau }=\frac{%
1}{\tau _{ex}}+\frac{1}{\tau _{0}}}$\cite{Haus}. Since the Raman effect will
excite both eigenmodes of the cavity (co- and counter-directionally
propagating modes), equal amplitude emission occurs along both waveguide
directions given by ${|s_{r}|^{2}=\frac{|E_{R}|^{2}}{2{\tau _{ex}}}}$. The
Raman intra-cavity gain coefficient is denoted as ${g_{R}^{c}}$, which can
be related to the more commonly used gain coefficient ${g_{R}}$(measured in
units of m/Watt) by, 
\begin{equation}
g_{R}^{c}\equiv \frac{c^{2}}{2n^{2}}\frac{1}{V_{eff}}g_{R}\text{ and }%
V_{eff}=\frac{\int \left\vert \vec{E}_{P}\right\vert ^{2}dV\int \left\vert 
\vec{E}_{R}\right\vert ^{2}dV}{\int \left\vert \vec{E}_{P}\right\vert
^{2}\left\vert \vec{E}_{R}\right\vert ^{2}dV}
\end{equation}%
where $V_{eff}$ is the effective modal volume \cite{Boyd}, and $\vec{E}$ is
the electric field vector \cite{VernooyQED}. The effective mode volume
accounts for the intensity dependent gain, and for silica microspheres and
microtoroids has approximately twice\ the value than the energy related
definition of mode volume. Steady state analysis of the coupled mode
equations, results in a clamped, cavity, pump field above threshold. This
clamping alters the coupling of pump power to the resonator, and, in turn,
the pump power dependence of Raman laser power such that the following
square root dependence results. 
\begin{equation}
P_{R}=\frac{\omega _{r}}{\omega _{p}}\left( \frac{1}{\tau _{ex}}\right)
^{2}\left( \frac{1}{2\tau _{0}}+\frac{1}{2\tau _{ex}}\right) ^{-2}\cdot
P_{t}\left( \sqrt{\frac{P}{P_{t}}}-1\right)
\end{equation}%
The physical origin of this square root dependence of the pump-to-Raman
conversion can be viewed as a \textquotedblleft pumping
inefficiency\textquotedblright\ i.e. the coupled pump power does not
increase linearly with launched fiber power. The nonlinear dependence of
coupled pump power can be illustrated by considering a pump wave that is
initially \textit{critically coupled \ }to the resonator. As noted above,
critical coupling features complete transfer and dissipation of power from
the resonator (i.e., zero transmission). In terms of the fields involved in
coupling both to and from the resonator, critical coupling results from the
destructive interference of the cavity leakage field with the transmitted,
pump field (i.e., the portion that does not couple to the resonator from the
waveguide). Once the onset of Raman lasing is reached, the cavity pump field
is clamped at the threshold value resulting in a fixed cavity pump leakage
field. Subsequent increase in launched pump power will imbalance the leakage
and the transmitted pump fields, giving rise to finite transmission and a
shift away from the critical point. The pump coupling to the resonator is
thereby less and less efficient as the pump field is increased. The
expression for the Raman threshold pump power can be factorized into terms
involving modal volume, waveguide-cavity coupling strength and cavity
lifetime (or Quality factor). To facilitate separation of the coupling and
intrinsic lifetime dependence, we use the dimensionless normalized coupling
parameter ${K_{p}=\left( \frac{\tau _{0}}{\tau _{ex}}\right) _{p}}$. In the
ideal case of a single-mode waveguide coupled to a whispering-gallery-mode
the waveguide transmission as a function of coupling is given by ${T=\left( 
\frac{1-K}{1+K}\right) ^{2}}$and ${K}$ typically varies exponentially with
the waveguide-microcavity "coupling gap" distance \cite{Haus}. Using these
definitions\ and under the assumption of equal coupling properties and
photon lifetimes for both the pump and Raman mode, i.e. ${K_{p}=K_{R}\equiv K%
}$ and $\tau _{R}=\tau _{p}$ ,the threshold expression is given by:%
\begin{equation}
P_{t}=C(\Gamma )\frac{\pi ^{2}n^{2}}{g_{R}\lambda _{p}\lambda _{R}}%
V_{eff}\cdot \left( \frac{1}{Q_{0}}\right) ^{2}\cdot \frac{\left( 1+K\right)
^{3}}{K}
\end{equation}%
Here we have also introduced ${C(\Gamma )}$ which is a possible correction
factor to account for intermode coupling of the degenerate clockwise and
counterclockwise propagating whispering-gallery modes. This factor will be
explained in the next section. The threshold expression follows an inverse
square dependence on the quality factor. This reflects the fact that an
increase in Q will cause a twofold benefit in terms of both reducing cavity
round trip losses that must be overcome for threshold as well as increasing
the Raman gain, due to the intensity dependence of the Raman gain
coefficient on the pump field. In addition, the equation shows that the
threshold scales linearly with the modal volume. Both the coupling and mode
volume dependence of the Raman threshold are examined experimentally in the
next section. When analyzing the coupling dependence under the assumption of
equal Raman and pump quality factors and coupling factors, the minimum
threshold occurs when ${Q_{ex}^{\min }=2\cdot Q_{0}}$ or ${K}^{\min }{=\frac{%
1}{2}}$, i.e. in the undercoupled regime with finite waveguide transmission
of ${T}^{\min }{=\frac{1}{9}(\symbol{126}11\%)}$. This minimum pump
threshold is given by,%
\begin{equation}
P_{t}^{\min }=C(\Gamma )\frac{\pi ^{2}n^{2}}{g_{R}\lambda _{p}\lambda _{R}}%
V_{eff}\cdot \left( \frac{1}{Q_{0}}\right) ^{2}\cdot \frac{27}{4}
\end{equation}%
It is worth noting that at this coupling condition, the circulating
pump-power in the resonator is not maximum. This can be understood since
minimum threshold represents an optimal balance of both pump coupling and
Raman mode coupling loss. The conversion of pump power to Raman power above
threshold can be characterized by the differential slope efficiency. The {%
bidirectional \ }external differential slope efficiency $\eta _{ex}$ is
derived by linearizing the expression for $P_{R}$ near the threshold
condition and is given by, 
\begin{equation}
\eta _{ex}\equiv \frac{{d}P_{Raman}}{{d}P_{launched}}=2\cdot \frac{\omega
_{R}}{\omega _{p}}\left( 1+\frac{1}{K}\right) ^{-2}
\end{equation}%
\begin{figure}[tbp]
{\centerline{%
\includegraphics[width=7.6cm]{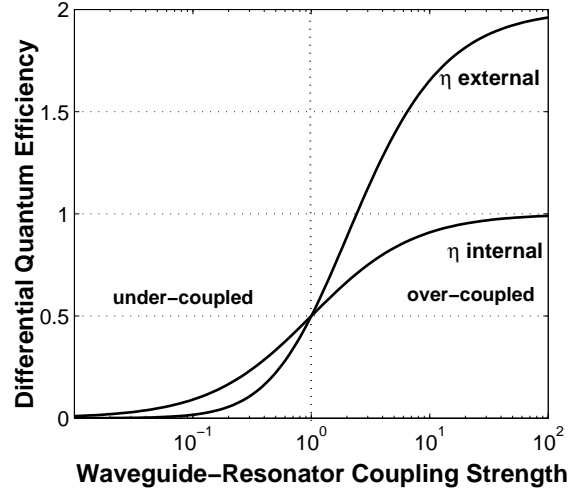}}
}
\caption{External and internal pump to Raman differential conversion
efficiency. The external efficiency refers to conversion of launched pump
power to Raman output power, while the internal efficiency is the coupled
pump power to Raman output power. As evident, the external efficiency can
exceed unity in the overcoupled regime}
\end{figure}
Figure 3 shows the differential slope efficiency as a function of coupling
strength. It is noteworthy that it approaches the value of ${2}\frac{\omega
_{R}}{\omega _{p}}$in the limit of strong overcoupling ${(\frac{\tau _{0}}{%
\tau _{ex}}=\infty )}$. Surprisingly, this value exceeds unity, indicating
that every waveguide pump photon added above threshold, is converted to more
than one Raman photon. This result can be understood by again considering \
the nonlinear dependence of coupled pump power, except this time in the
over-coupled regime. In particular, the differential increase in coupled
pump power grows more quickly in the overcoupled (more slowly in the
undercoupled regime) than the differential increase in launched pump power.
This leads to the interesting effect that the differential photon conversion
efficiency can exceed unity. Taking into account the nonlinear dependence of
coupled pump power by defining the internal differential efficiency ${\eta
_{int}}$ as the coupled (as distinct from the launched) pump-to-Raman power,
the efficiency approaches, as expected , the value $\frac{\omega _{R}}{%
\omega _{p}}$in the limit of strong overcoupling.%
\begin{equation}
\eta _{int}\equiv \frac{{d}P_{Raman}}{{d}P_{coupled}}=\frac{\omega _{R}}{%
\omega _{p}}\left( 1+\frac{1}{K}\right) ^{-1}
\end{equation}%
Figure 3 shows both the internal and external differential Raman conversion
efficiencies as a function of coupling strength.

\subsection{The effect of intermode coupling on stimulated Raman scattering}

In the ultra-high-Q regime, resonances of a whispering-gallery-type
microcavity are often split into doublets\cite{Braginsky} \cite{Weiss}. This
splitting is due to coupling of the degenerate clockwise and
counterclockwise propagating modes by either intrinsic or surface scattering
centers. The modified coupling properties have been extensively studied \cite%
{KippenbergOL}. Here we briefly summarize the results of this study and
analyze the effect of intermode coupling on Raman scattering. The extent to
which intermode coupling modifies the waveguide coupling properties can be
described by the dimensionless intermode coupling parameter $\Gamma =\frac{%
\tau _{0}}{\gamma }$, where $1/{\gamma }$ is the rate of coupling of the
degenerate clockwise and counterclockwise modes. It is easily measured as
the linewidth normalized splitting in the undercoupled regime\ (see figure 4
inset). The presence of intermode coupling has several consequences. First,
the critical point (as defined by vanishing waveguide transmission) is
shifted towards a coupling point, which, using the conventional coupling
terminology, is considered overcoupled as $\tau _{ex}<\tau _{0.}$The point
of vanishing transmission occurs at: 
\begin{equation}
K_{crit}=\sqrt{\Gamma ^{2}+1}
\end{equation}%
Second, the shifted critical point is accompanied by a maximum reflection
into the backwards direction of the waveguide (i.e. the contra-directional
waveguide mode). The magnitude of the reflection at the above modified
critical point is given by%
\begin{equation}
R^{\max }=\left( \frac{\Gamma }{\sqrt{\Gamma ^{2}+1}+1}\right)
\end{equation}%
Third, the leakage of the cavity field into the backwards waveguide
direction causes a reduction of the cavity buildup factor with \ respect to
the ideal case in the absence of intermode coupling. In the limit of strong
modal coupling, the cavity buildup factor is reduced by a factor of $2$,
which subsequently causes a twofold increase in the threshold necessary to
achieve Raman lasing. Figure 4 shows the circulating power correction factor 
${C(\Gamma })$ as a function of the dimensionless intermode coupling
parameter ${\Gamma .}$ In the presence of modal coupling the waveguide
coupling condition for minimum threshold experiences a slight shift towards
overcoupling with the maximum shift occurring at ${\Gamma \approx 1.52}$. In
the regime of very strong modal coupling the condition of minimum Raman
threshold approaches again the original condition ${K=1/2}$. 
\begin{figure}[tbp]
{\centerline{%
\includegraphics[width=7.6cm]{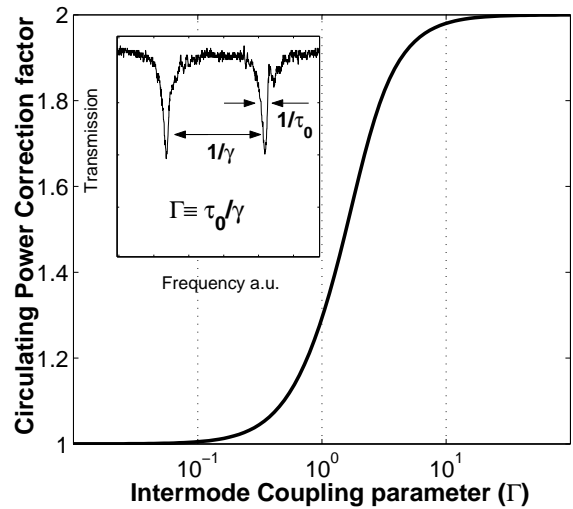}}}
\caption{Power correction factor (which is the ratio defined by maximum
circulating power in the absence and presence of modal coupling) as a
function of $\Gamma $-parameter. The inset shows a frequency scan of a
doublet of an ultra-high-Q mode.}
\end{figure}

\subsection{Analysis of Cascaded Raman Scattering in high-Q microcavities}

The first Raman field can itself act as a secondary pump field and generate
further Raman modes. This process of cascaded Raman scattering can be
described by including higher order coupling terms into the coupled mode
equations of pump and Raman fields as shown below\cite{Min}. 
\begin{eqnarray}
\frac{dE_{p}}{dt} &=&\left[ -\left( \frac{1}{2\tau _{t}}\right)
_{P}-g_{_{R1}}^{c}\left( \frac{\omega _{_{P}}}{\omega _{_{R}}}\right)
\left\vert E_{R1}\right\vert ^{2}\right] E_{p}+\sqrt{\frac{1}{\tau _{ex}}}s
\\
\frac{dE_{_{R1}}}{dt} &=&\left[ -\left( \frac{1}{2\tau _{t}}\right)
_{_{R1}}+g_{_{R1}}^{c}\left\vert E_{p}\right\vert ^{2}-g_{_{R2}}^{c}\left( 
\frac{\omega _{_{R1}}}{\omega _{_{R2}}}\right) \left\vert E_{R1}\right\vert
^{2}\right] E_{_{R1}}  \notag \\
\frac{dE_{_{R2}}}{dt} &=&\left[ -\left( \frac{1}{2\tau _{t}}\right)
_{_{R2}}+g_{_{R2}}^{c}\left\vert E_{R1}\right\vert ^{2}-g_{_{R3}}^{c}\left( 
\frac{\omega _{_{R2}}}{\omega _{_{R3}}}\right) \left\vert E_{R3}\right\vert
^{2}\right] E_{_{R2}}  \notag \\
&&.......  \notag \\
\frac{dE_{_{RN}}}{dt} &=&\left[ -\left( \frac{1}{2\tau _{t}}\right)
_{_{RN}}+g_{_{RN}}^{c}\left\vert E_{R(N-1)}\right\vert ^{2}\right] E_{_{RN}}
\notag
\end{eqnarray}%
where $N$ is the Raman order. To find the corresponding thresholds and
output powers for these higher order processes, the set of equations can be
solved iteratively in steady state. Here, we have introduced the
dimensionless coefficients $c_{i}$. 
\begin{equation}
c_{i}\equiv \frac{\omega _{i}}{\omega _{i+1}}\cdot \frac{g_{i+1}^{c}}{%
g_{i+2}^{c}}=\frac{\omega _{i}}{\omega _{i+1}}\cdot \frac{V_{eff}(\lambda
_{i+2})}{V_{eff}(\lambda _{i})}
\end{equation}%
The general solutions for the threshold of the even and odd order $N^{th}$
Raman modes are given by the following expressions. As in the previous
section, we have assumed equal coupling strengths and intrinsic Q factors
for the pump and Raman modes. 
\begin{eqnarray}
P_{t}^{N=2m} &=&\frac{1}{g_{R}^{c}}\frac{\tau _{ex}}{\left( \tau _{t}\right)
^{3}}\left( \sum_{i=0}^{m}\left( c_{i}\right) ^{i}\right) ^{2}\left(
\sum_{i=0}^{m-1}\left( c_{i}\right) ^{i}\right)  \notag \\
P_{t}^{N=2m+1} &=&\frac{1}{g_{R}^{c}}\frac{\tau _{ex}}{\left( \tau
_{t}\right) ^{3}}\left( \sum_{i=0}^{m}\left( c_{i}\right) ^{i}\right) ^{3}
\end{eqnarray}%
As evident from these expressions even and odd order stokes fields exhibit
different threshold powers as a function of stokes order (N). When
considering Raman scattering in silica at optical frequencies, one can
approximate the above expressions by taking $c_{i}\approx 1$ since the Raman
shift is small compared to the optical frequency. In addition it is assumed
that the mode volume is wavelength independent. Under this assumption, the
threshold expressions reduce to: 
\begin{eqnarray}
P_{t}^{N=2m+1} &=&C(\Gamma )\frac{\pi ^{2}n^{2}}{g_{R}\lambda _{p}\lambda
_{R}}V_{eff}\frac{1}{Q_{0}^{2}}\frac{\left( 1+K\right) ^{3}}{K}\cdot \frac{%
(N+1)^{3}}{8} \\
P_{t}^{N=2m} &=&C(\Gamma )\frac{\pi ^{2}n^{2}}{g_{R}\lambda _{p}\lambda _{R}}%
V_{eff}\frac{1}{Q_{0}^{2}}\frac{\left( 1+K\right) ^{3}}{K}\cdot \frac{%
N(N+2)^{2}}{8}  \notag
\end{eqnarray}%
It follows that the threshold for cascaded Raman oscillation exhibits a $%
cubic$ dependence on Raman order N. The emission power dependences vary
depending upon whether the highest order wave is even or odd. For the odd
order case, all odd orders increase as the square root of the pump power and
even orders are clamped. For the even order case, all even order lines
increase linearly with pump power while odd orders are clamped. Figure 5
illustrates this behavior showing the Raman output for several stokes orders
as a function of input pump power. The analytic expressions for the Raman
output power in these cases are given by: 
\begin{eqnarray}
P^{N=2m+1} &=&\eta _{ex}^{N}\cdot 2\left( \sqrt{P_{t}^{N}P}-P_{t}^{N}\right)
\\
P^{N=2m} &=&\eta _{ex}^{N}\cdot \left( P-P_{t}^{N}\right)  \notag
\end{eqnarray}%
The differential power conversion efficiencies can be obtained by
linearizing the above expressions near the threshold condition. The external
and internal differential efficiencies decrease steadily as a function of
stokes order ($N$). For optical frequencies that are much larger than the
Raman shift, the external differential conversion efficiency reduces to:%
\begin{eqnarray}
\eta _{ex}^{N=2m} &=&\frac{\lambda _{p}}{\lambda _{RN}}\left( 1+\frac{1}{K}%
\right) ^{-2}\cdot \frac{16}{\left( N+2\right) ^{2}} \\
\eta _{ex}^{N=2m+1} &=&\frac{\lambda _{p}}{\lambda _{RN}}\left( 1+\frac{1}{K}%
\right) ^{-2}\cdot \frac{8}{\left( N+1\right) ^{2}}  \notag
\end{eqnarray}%
\begin{figure}[tbp]
{\centerline{\includegraphics[width=7.6cm]{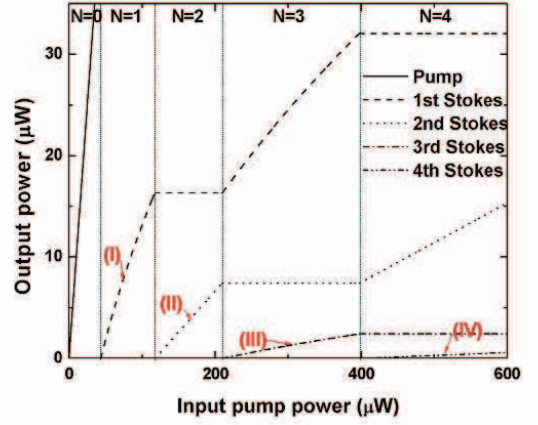}}}
\caption{Theoretical plot of cascaded Raman output power versus pump power
for up to fourth order Stokes power. Even and odd ordered Raman modes
exhibit alternating clamping behaviour.}
\end{figure}
For high order (N) Raman fields, the external differential conversion
efficiency thus follows a $1/N^{2}$-dependence.

\section{Experimental study of Stimulated Raman Scattering in ultra-high-Q
microspheres}

We have observed stimulated Raman scattering using fiber-taper-coupled,
ultra-high-Q, surface-tension-induced silica microspheres. Tapered optical
fibers provide both a very efficient and practical means for pumping and
laser signal extraction through the same fiber. The experimental setup in
this study is identical to the one reported earlier\cite{SpillaneNature} and
the tapered fiber was attached to a piezoelectric stage which allowed
precise control of the taper-microcavity coupling gap (20 nm resolution) and
variation of the waveguide-resonator coupling strength. The Q-factors
obtained in the microspheres were typically in the range of $1-2\times
10^{8} $ and taper insertion loss (fiber to fiber) was usually less than
5\%. \ 

\begin{figure}[tbp]
{%
\centerline{
\includegraphics[width=7.6cm]{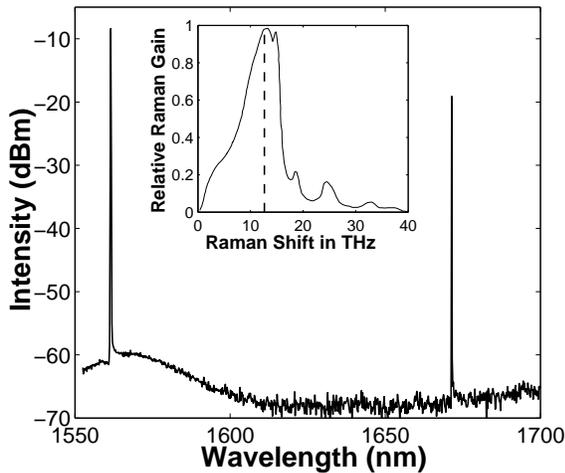}}}
\caption{Single mode Raman lasing in a UHQ spherical microcavity. The pump
wavelength is located at 1550 nm and Raman lasing appears at 1660 nm. The
inset shows the Raman frequency shift (designated by the dotted line)
overlaid with the Silica Raman gain spectrum.}
\end{figure}
\begin{figure}[tbp]
{\centerline{%
\includegraphics[width=7.6cm]{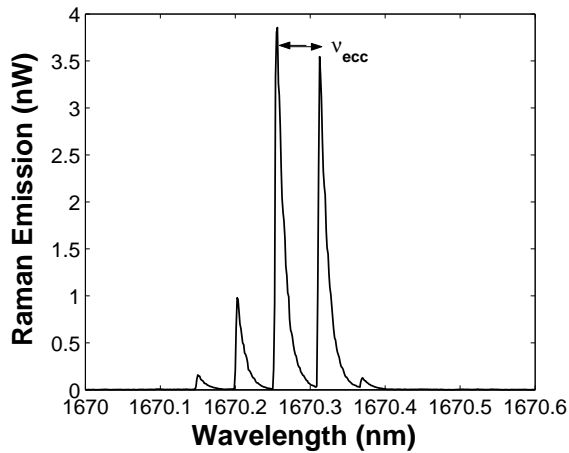}}}
\caption{Spectrally resolved Raman emission of the ultra-high-Q microsphere
shown in the previous figure, revealing that 5 eccentricity split azimuthal
modes were oscillating simultaneously. The pump power in this measurement
was adjusted to slightly above Raman oscillation threshold. The assymmetric
line-shape of the Raman modes is caused by repeatedly scanning the pump
laser through the pump resonance, which causes a redshift of the cavity
resonances, due to heating and subsequent expansion of the cavity (thermal
bistability as noted in \protect\cite{Braginsky}).}
\end{figure}
To excite the UHQ modes, we used a narrow-linewidth, external-cavity laser
emitting in the 1550 nm band. In the case of microspheres, the WGM field
spatial forms are known analytically and characterized by the radial,
angular, azimuthal and polarization mode number $(n,\ell ,m,p)$. Due to
fabrication-induced eccentricity, the $(2\ell +1)$-fold-degeneracy of the
azimuthal modes is lifted yielding a complex mode spectrum. For the spheres
considered in this work the eccentricity-induced splitting was in the range
of several GigaHertz. Microtoroids on-a-chip, on the other hand, possess a
significantly reduced mode spectrum due to the strong azimuthal modal
confinement provided by the toroid geometry. This both simplifies their
spectra and enables operation of microtoroid-Raman lasers in the desirable
single mode regime, which is of significant practical importance.

Stimulated Raman oscillation was observed by pumping a single WGM and
monitoring the transmission using an optical spectrum analyzer. Once the
threshold for SRS was exceeded, lasing modes in the 1650-nm band could be
observed, in correspondence with the peak Raman gain which occurs
downshifted in frequency by approximately 14 THz relative to the pump
frequency (wavelength shift of approximately 110 nm).\ Figure 6 shows Raman
emission for an ultra-high-Q microsphere. The Raman emission with respect to
the gain peak is provided in the inset of figure 6. Since the fundamental
whispering gallery modes $(n=1,\ell =m,p=TM)$ are most tightly confined
(i.e., smallest mode volume), Raman lasing is expected to occur first for
these modes. The presence of nearly degenerate azimuthal modes in a
spherical microcavity (i.e., weak eccentricity splitting), causes
simultaneous oscillation on several azimuthal modes. Figure 7 shows a higher
resolution spectral scan of the spectrum in figure 6. Several azimuthal
modes can be observed to be oscillating simultaneously.

The threshold formula predicts a strong dependence of the Raman threshold on
waveguide coupling. Figure 8 shows the measured Raman threshold as a
function of taper-microcavity gap distance for a fundamental WGM. 
\begin{figure}[tbp]
{\centerline{\includegraphics[width=7.6cm]{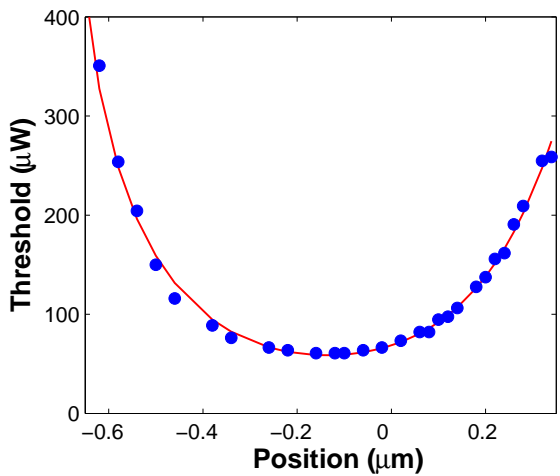}}}
\caption{The dependence of Raman threshold on loading as evidenced by the
taper-micro-cavity gap distance. The taper-microcavity gap is measured in
microns with respect to the critical point (defined as vanishing taper
transmission), and positive separation corresponds to overcoupling. The
minimum threshold is 62 $\protect\mu $W of launched pump power (measured
before the taper which includes the insertion loss of the optical fiber
taper). Minimum threshold occurs ca. 120 nm undercoupled (corresponding to a
transmission of 13\%).}
\end{figure}
The solid line is a fit using the theoretical results from the first section
(equation 3) and assuming equal pump and Raman Q factors and coupling
conditions. The minimum threshold does indeed occur undercoupled with finite
taper transmission at ${T=13\%}$ in good agreement with the theoretically
predicted value of ${T=1/9}$.

As a further verification of the threshold formula we compared the
theoretical minimum threshold value with the observed value. The quality
factor and the mode splitting of the whispering-gallery mode were measured
by performing a \ linewidth sweep in the undercoupled regime, where the
backscattering-induced doublet structure is most pronounced. These
measurements yielded $Q_{0}=1\times 10^{8}$ and $\Gamma =2$ . The size of
the microsphere was inferred from the free spectral range, $\Delta \lambda
=10.5$ $nm$ i.e. $50$ $\mu m$ diameter, where the free-spectral range
denotes here modes with successive angular mode number ${\ell }$. The mode
volume was calculated using analytic expressions based on estimated mode
numbers for the fundamental WGM ${(n=1,p=TM,m=\ell ,\ell \approx 139)}$.
Calculations\cite{Buck} \cite{LittleHaus} yielded a modal volume of ca.$1300$
${\mu m^{3}}$. The overlap factor for pump and Raman mode is assumed to be
unity. Using these values, the theoretically expected minimum threshold is
given by $50$ ${\mu Watts}$ which is in good agreement with the
experimentally measured value of 62 ${\mu Watts.}$

\begin{figure}[tbp]
\centerline{%
\includegraphics[width=7.6cm]{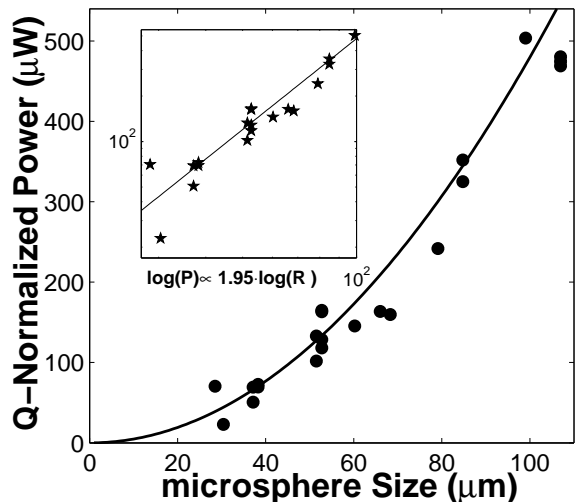}}
\caption{The Raman oscillation pump threshold of a spherical UHQ microcavity
as a function microsphere diameter. \ To compare microspheres with different
resonance characteristics, the Raman threshold was Q-normalized to $%
Q=1.0\times 10^{8}$ and modal coupling normalized to $\Gamma =0$. The inset
shows a double logarithmic plot of the data. A linear fit obtains a
dependence of \ $R^{1.95}$ which is in close agreement with the
theoretically expected dependence of $R^{1.83}$.}
\end{figure}

The dependence of the Raman threshold on the modal volume was also
investigated. For this measurement the Raman threshold was measured for
micro-sphere resonators having varying diameters in the range of ca. 25-120 $%
\mu m$. (For diameters smaller than 25-$\mu m$ thermally induced wavelength
shifts lead to pumping instabilities.) In this diameter range, the mode
volume follows an approximately quadratic dependence (actual inferred
exponent is ${V\propto R}^{1.83}$ \cite{LittleHaus} \cite{Buck}) on the
sphere radius. As an aside, for smaller spheres the mode volume deviates
from this behavior and ultimately, for very small diameters, the mode volume
increases due to weakening of the whispering gallery confinement\cite{Buck}.
The minimum mode volume occurs for a radius of $6.9\mu m$ (for $\ell =m=34$)%
\cite{Buck} for 1550 nm wavelength and the mode volume is $V_{\min
}=173.1\mu m^{3}$. However, this size is not optimum for stimulated Raman
scattering as the additional benefit of reduced mode volume is more than
offset by the significant decrease in Q factor to ${10^{5}}$\cite{LittleHaus}%
(Threshold power ${\propto V/Q^{2}}$). For small mode volumes, it has been
predicted that the gain coefficient can exhibit a dependence on mode volume
due to cavity QED effects\cite{CampilloQED}. \ However, in the case of
stimulated Raman Scattering in silica microspheres these effects are not
expected to be observable \cite{Matsko} and an approximately quadratic
relationship due to the mode volume is predicted.

Figure 9 shows the experimental results for threshold versus microsphere
diameter. In this experiment, the minimum Raman threshold ${P_{t}^{\min }}$,
the microsphere size (as inferred from the free-spectral-range), the
intrinsic Q in the pump band (${Q_{0}}$) and the intermode coupling
parameter (${\Gamma }$) were measured. To extract the volumetric dependence
of the Raman threshold using data from cavities having different resonant
characteristics, the threshold data were normalized to the set of parameters 
${(Q_{0}=10^{8},\Gamma =0,g_{R}=g_{R}^{\max })}$. The result of this
procedure is shown in figure 9. The data indeed show a quadratic dependence
on R (the actual fitted exponential from a double logarithmic plot is 1.95
and is in good agreement with the expected value of 1.83) and confirm the
linear relationship of the Raman threshold on the mode volume as predicted
by equation 4.

\subsection{Cascaded Raman Scattering in ultra-high-Q microspheres}

\bigskip 
\begin{figure}[tbp]
\centerline{\includegraphics[width=8cm]{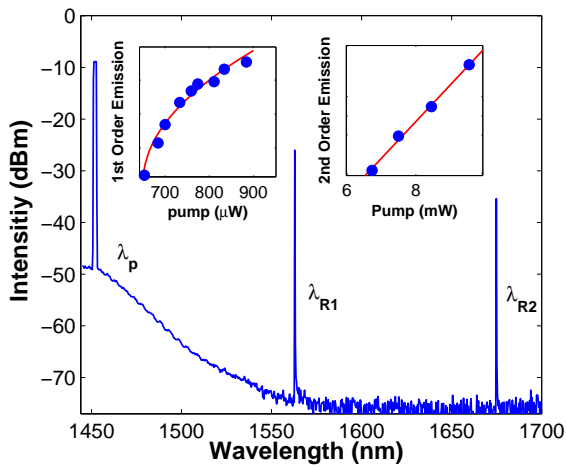}}
\caption{Cascaded Raman Scattering in a 58-$\protect\mu m-$diameter
microcavity.\ The pump WGM is located at around 1450 nm. The insets show the
pump-to-Raman conversion for first\ (left inset) and second order (right
inset) Raman modes (measured on different microcavities). The $1^{st}$ order
Raman mode exhibits a square-root, and the $2^{nd}$ order Raman mode a
linear pump-to-Raman conversion characteristic, in agreement with the
theoretical prediction. Solid lines: A theoretical fit using equations 12.}
\end{figure}

\begin{figure}[tbp]
\centerline{\includegraphics[width=7cm]{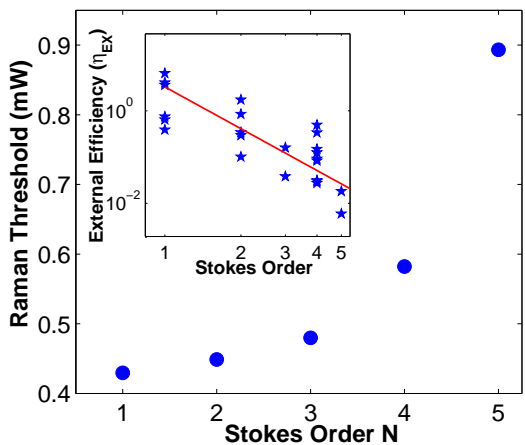}}
\caption{5$^{th}$ Order Cascaded Raman Scattering. The threshold of the
cascades are plotted as a function of Stokes order (N). The data was
acquired on a single microsphere of 50-$\protect\mu m-$diameter excited
using a WGM at 980-nm wavelength.The inset shows the efficiency of the
cascades as a function of Cascading order, measured on several devices..}
\end{figure}

In addition to first order Raman scattering, we have also observed cascaded
Raman scattering. Figure 10 shows a typical cascaded Raman spectrum, for a
UHQ\ microcavity pumped at 1450 nm. The pump-to-Raman conversion
characteristics for first order Raman scattering and the 2$^{nd}$ order
Raman mode are shown in the inset. It can be seen that the first order mode
does indeed exhibit a square-root dependence on the launched pump power.\
The solid line is a fit using equation 12. The higher order Raman mode
exhibits the expected linear increase with pump power.

To study cascaded Raman scattering beyond 2$^{nd}$ order, experiments using
a 980 nm wavelength pump \cite{Min} were employed. The shorter wavelength
pump allowed the observation of up-to 5$^{th}$-order cascades (from 980-1300
nm) owing to the reduced mode volume at shorter wavelengths, and the higher
Raman gain coefficient $(g_{R}\propto 1/\lambda )$. With less than 900 $\mu $%
-Watts of launched fiber power up to fifth order Stimulated Raman Scattering
was observed \cite{Min}, and the threshold and efficiency of the cascades
measured. Figure 11 shows the Cascaded Raman threshold for a microsphere as
a function of the order of the cascade, for fixed coupling condition. The
solid line is a cubic fit (as predicted by equation 11), which yields good
qualitative agreement with the experimentally measured thresholds. The inset
of figure 11 shows a double logarithmic plot of the measured differential
conversion efficiency of the cascaded Raman scattering process.\ The
efficiency of the cascades decreases as a function of Stokes order, as is
theoretically predicted by equation 13.\ 

\section{Raman Scattering in Ultra-high-Q toroid microcavities on-a-chip.}

\bigskip

Toroid microcavities allow on-chip integration of ultra-high-Q performance
with other optical, mechanical and electrical functionality.\ In addition
the wafer-scale fabrication process allows precise dimensional control and
parallelism. Figure 12 shows a scanning electron microscope of a toroid
microcavity. In addition to the design freedom and control brought about by
micro-fabrication and integration possibilities, toroid microcavities also
exhibit significant advantages in terms of their Raman-emission properties.
Whereas microsphere cavities show low threshold operation, their emission is
inherently multi-mode due to the presence of azimuthal modes, as shown in
figure 7 of the previous section. In contrast, toroid microcavities
on-a-chip have a significantly reduced mode spectrum, such that \textit{%
single mode} Raman lasing can be observed. The reduced number of azimuthal
modes is due to the cavity geometry, which supports only a few modes in the
azimuthal direction due to the toroidal confinement\cite{IlchenkoMicrotorus}%
. Figure 13 shows the mode spectrum of a toroid microcavity. Only two higher
order modes are present in the spectrum, and successive modes are separated
by the free spectral range of the cavity, which in the figure is 10 nm. This
is in contrast to a microsphere of identical principal diameter which would
support $2\ell +1$ azimuthal modes weakly split by eccentricity.  
\begin{figure}[tbp]
\centerline{\includegraphics[width=6cm,
height=5cm]{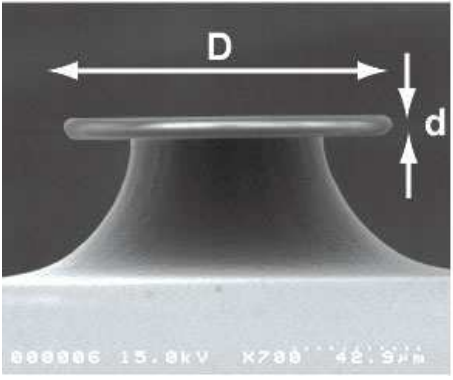}}
\caption{Scanning-electron-micrograph of a toroid microcavity. The principal
diameter (D) and minor toroid diameter(d) are shown.}
\end{figure}

\begin{figure}[tbp]
\centerline{\includegraphics[width=7.6cm]{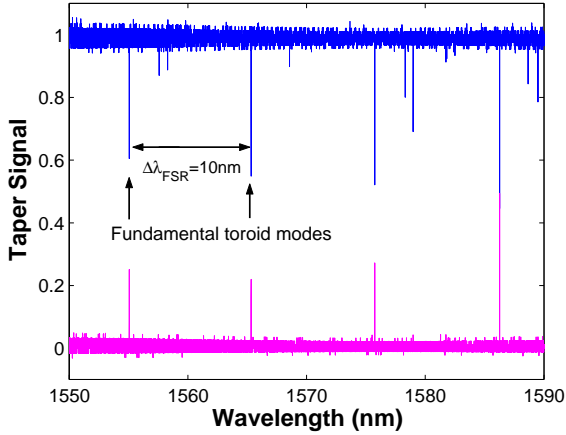}}
\caption{The recorded mode spectrum (both transmission and reflection) of a
50-$\protect\mu m-$diameter toroid microcavity, exhibiting a strongly
reduced mode spectrum. The reduced density of modes is due to the encreased
azimuthal confinement, and is similar to spectra obtained with spheroids%
\protect\cite{IlchenkoMicrotorus}.}
\end{figure}

The suppression of azimuthal modes, has important consequences on the
spectral emission properties of single and cascaded Raman scattering. Most
notably, single transverse mode Raman oscillation can be observed over a
large range of pump powers. Figure 14 shows a toroid microcavity Raman
laser, which exhibits single mode oscillation at emission power levels up to
160 $\mu Watts$ with high efficiency (45\% at the critical point). The
single-mode emission property is a significant advantage in practical
applications of nonlinear optical oscillators. 
\begin{figure}[tbp]
\centerline{%
\includegraphics[width=7.6cm]{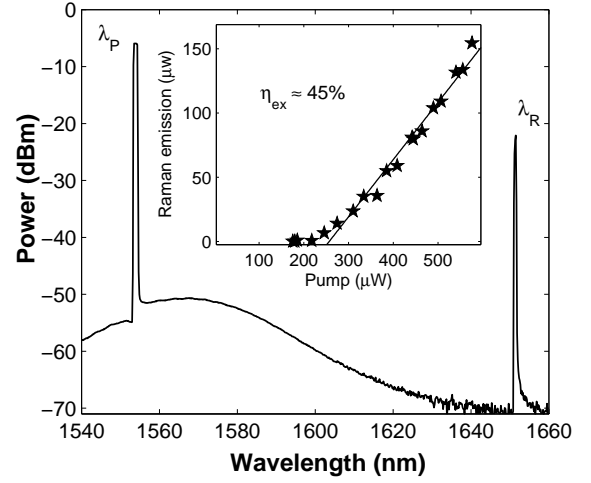}}
\caption{Single mode Raman emission of a ca. 60-$\protect\mu m-$diameter
toroid microcavity.The pump is located at 1550 nm.}
\end{figure}
Toroidal and spherical microcavities having similar Q and outer diameter
will differ in modal volume. Toroids, owing to the added transverse
confinement will have a reduced mode volume and hence a lower Raman
threshold. The degree to which the threshold is reduced will depend on the
"aspect ratio" of the toroid or D/d where D is the outer or principal
diameter and d is the minor diameter (see figure 12). The modal volume of
the fundamental toroid mode in a D= 50 $\mu m$ toroid plotted versus $d$ is
provided in figure 15. The calculation uses a finite element mode solver as
there are no analytical expressions available for toroid mode volume. The
case of d = 50 $\mu m$ corresponds to a sphere. Overall, the toroidal mode
volume can be seen to be lower than the sphere. There are also two distinct
regimes of mode volume behavior. In the first, the mode volume reduces very
slowly as d is reduced ($V\propto d^{1/4}$). This regime features a weak
lateral confinement very similar in nature to that of the sphere and related
to the gentle transverse curvature of the toroidal dielectric boundary. In
the second regime, the mode volume reduces very quickly as d is reduced.
This regime is characterized by strong, lateral, index confinement resulting
from the toroidal boundary being comparable in diameter to the mode field.
The inset to this figure provides mode fields for the toroid in these two
regimes to further illustrate this idea. It is clear that substantial
reductions in threshold are possible using these devices. We are currently
characterizing the reductions possible experimentally.

\begin{figure}[tbp]
\centerline{%
\includegraphics[width=7.6cm]{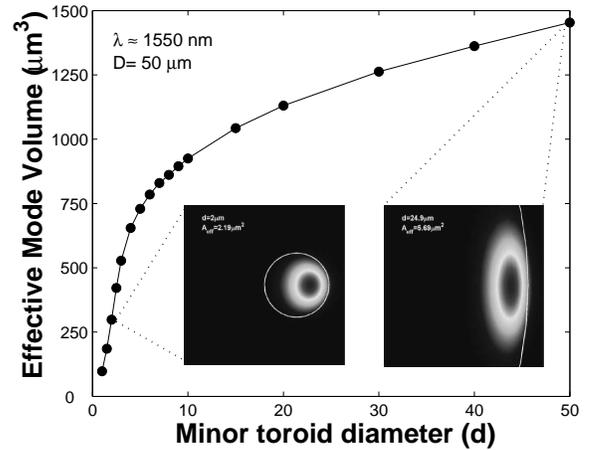}}
\caption{Numerically calculated effective mode volume of a toroid
microcavity as a function of the cross-sectional diameter of the toroid (and
fixed principal toroid diameter D=50$\protect\mu m$). The circles denote the
numerically calculated effective mode volume for a fundamental toroid WGM in
the 1550-nm band for (TE) polarization. The inset shows the modeled TE-mode
intensity distribution for a toroid with 2 $\protect\mu m$ and 50 $\protect%
\mu m$ toroid-cross sectional diameter (d) using the finite-elements method.}
\end{figure}

\section{Summary}

In summary we have experimentally and theoretically analyzed Raman
oscillation in fiber-taper-coupled microspheres and microtoroids on-a-chip.
A theoretical analysis was presented using the coupled mode equations for
the pump and Raman WGMs. Using these equations, the threshold condition for
stimulated Raman scattering was derived and the relative importance of
waveguide coupling strength, mode volume and intrinsic resonator Q were
described. These theoretical dependences were verified experimentally.
Furthermore the analysis was extended to the case of cascaded Raman
oscillation and threshold and efficiency expressions were derived for
higher-order Raman fields. This analysis revealed that odd and even order
Raman lines exhibit different pump-to-Raman emission characteristics. Even
order Stokes fields are found to exhibit a linear increase in generated
Raman power as a function of pump power, whereas odd-order Stokes fields
exhibit a square root dependence. Analysis showed and experiment confirmed
that the threshold for N-th-order cascaded Raman oscillation exhibits a
cubic dependence on order and that the associated efficiency of the process
scales inverse quadratically with order.

Microtoroids, in addition to having significant practical advantages with
respect to their chip-based fabrication, have both spectral and power
efficiency advantages in comparison to Raman oscillation in microspheres.
Their stronger lateral confinement provides two distinct benefits. First, a
drastic reduction in the complexity of the mode spectrum enabling
single-mode oscillation in the microtoroid based device. Second, a
controllable and reduced mode volume so that for comparable Q factors,
microtoroid devices should have lower threshold pump powers. We are
currently investigating the latter advantage experimentally.

The importance of fiber taper coupling in these measurements cannot be over
emphasized. These inherently fiber-compatible waveguides provide exceptional
coupling efficiencies to and from the ultra-high-Q devices. They are also
indispensable in coupling to the microtoroid devices which reside near the
surface of a silicon wafer. Using taper coupling, the lowest threshold
observed in this study was 62 $\mu $Watts of launched power, a value which
is nearly 3-orders of magnitude lower than for free-space illumination of
micro-droplets.

\section{\protect\bigskip Acknowledgement}

\bigskip This work was supported by the NSF, DARPA and the Caltech Lee
Center for Advanced Networking.

\end{document}